\documentclass[aip,jap,reprint,graphicx]{revtex4-1}
\usepackage{graphics}
\usepackage{dcolumn}
\usepackage{bm}
\usepackage{amsmath}
\usepackage{amssymb}
\usepackage{epsfig}
\usepackage{pictex}
\begin{document}
\title{ A site-selective antiferromagnetic ground state in layered pnictide-oxide
        BaTi$_{2}$As$_{2}$O }
\author{Xiang-Long Yu}
\affiliation{Key Laboratory of Materials Physics,
             Institute of Solid State Physics, Chinese Academy of Sciences,
             P. O. Box 1129, Hefei 230031, China}
\affiliation{University of Chinese Academy of Sciences,
             Beijing 100000, China}
\author{Da-Yong Liu}
\affiliation{Key Laboratory of Materials Physics,
             Institute of Solid State Physics, Chinese Academy of Sciences,
             P. O. Box 1129, Hefei 230031, China}
\author{Ya-Min Quan}
\affiliation{Key Laboratory of Materials Physics,
             Institute of Solid State Physics, Chinese Academy of Sciences,
             P. O. Box 1129, Hefei 230031, China}
\author{Ting Jia}
\affiliation{Key Laboratory of Materials Physics,
             Institute of Solid State Physics, Chinese Academy of Sciences,
             P. O. Box 1129, Hefei 230031, China}
\author{Hai-Qing Lin}
\affiliation{Beijing Computational Science Research Center, Beijing 100084, China}
\author{Liang-Jian Zou}
\email[Author to whom correspondence should be addressed. Electronic mail: ]{zou@theory.issp.ac.cn}
\affiliation{Key Laboratory of Materials Physics,
             Institute of Solid State Physics, Chinese Academy of Sciences,
             P. O. Box 1129, Hefei 230031, China}
date{\today}

\begin{abstract}
The electronic and magnetic properties of BaTi$_{2}$As$_{2}$O have been
investigated using both the first-principles and analytical methods.
The full-potential linearized augmented plane-wave calculations
show that the most stable state is a site-selective
antiferromagnetic (AFM) metal with a $\text{2}\times \text{1}\times
\text{1}$ magnetic unit cell containing two nonmagnetic Ti atoms and two
other Ti atoms with antiparallel moments. Further analysis to Fermi
surface and spin susceptibility shows that the site-selective AFM ground
state is driven by the Fermi surface nesting and the Coulomb correlation.
Meanwhile, the charge density distribution remains uniform, suggesting that the
phase transition at $200$ K in experiment is a spin-density-wave (SDW)
transition.
\end{abstract}

\pacs{31.15.A-, 75.10.Jm, 74.20Mn, 75.40.Cx}

\maketitle

\section{Introduction and Numerical Method}
Recently Wang $et\ al.$ synthesized a new layered pnictide oxide BaTi$_{2}$As$_{2}$O \cite{JPCM22.075702}. This compound is isostructural to
BaTi$_{2}$Sb$_{2}$O and Na$_{2}$Ti$_{2}$$Pn_{2}$O ($Pn=$As,Sb). These
four pnictide oxides exhibit similar anomalies in magnetic susceptibility and electrical resistivity at a certain temperature
$T_p$ \cite{JACS134.16520,JPSJ81.103706,ZAAC584.150}. This shows that a phase transition
occurs around $T_p$, very analogous to the prototype LaFeAsO\cite{JACS130.3296}.
The anomaly of BaTi$_{2}$As$_{2}$O at $200$ K is ascribed to a possible
spin-density-wave (SDW) or charge-density-wave (CDW) transition\cite{JPCM22.075702}.
However, the occurrence of a CDW state is usually accompanied by the distortion of
crystal lattice\cite{bookGG}. The distortion was not clearly observed in the
structure measurement. It is probable that the ground state is an SDW phase.
Nevertheless, its groundstate electronic and magnetic structures remain an open
question. In this paper we
study the electronic and magnetic properties of BaTi$_{2}$As$_{2}$O by
combining the first-principles electronic structure calculations and the analytical
method. Our study on BaTi$_{2}$As$_{2}$O will not only uncover
the underlying mechanism for its anomalous properties, but also address the
properties of the isostructural compounds BaTi$_{2}$Sb$_{2}$O and
Na$_{2}$Ti$_{2}$$Pn_{2}$O.

\begin{figure}[tp]
{\epsfig{figure=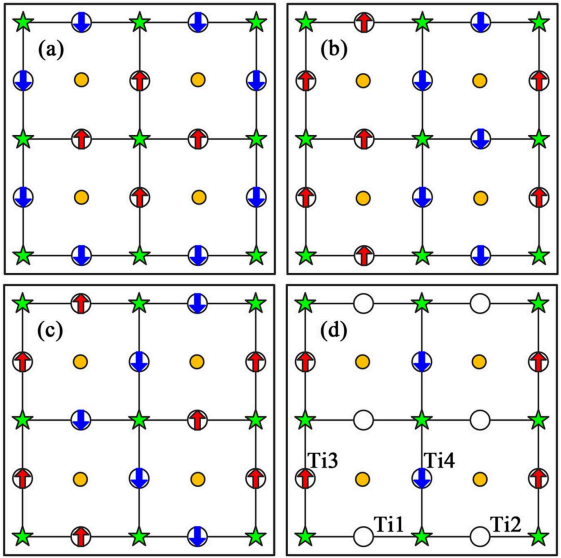,width=7.5cm,angle=0}}
\caption{(Color online) Four magnetic structures on Ti$_{2}$As$_{2}$O layer:
(a) BCAFM, (b) BAFM, (c) ZAFM and (d) site-selective AFM ordered phases.
The big and small circles represent Ti and O atoms, respectively; and the green
star represents As atom.}
\label{fig:AFMstruct}
\end{figure}

The electronic structure calculations are performed by using the full-potential
linearized augmented plane-wave (FPLAPW) code of the package WIEN$2$K.
The lattice parameters of BaTi$_{2}$As$_{2}$O ($a=4.045608{\text{{\AA}}}$,
$c=7.27228{\text{{\AA}}}$) for our calculations are provided by Wang $et\ al.$
\cite{JPCM22.075702}. The muffin-tin sphere radii are $2.5$, $2.02$, $2.4$ and
$1.79$ Bohr for Ba, Ti, As and O atoms, respectively.  The cut-off parameter
R$_{mt}$K$_{max}$ is chosen to be $7.0$. Our electronic structure calculations
are based on the Perdew-Barke-Ernzerhof generalized gradient approximation (GGA)
\cite{PRL77.3865} and its correlation correction (GGA+U).
The number of \textbf{k} points is taken $2000$ for unit cell and $1000$ for
supercell.
Referring to the realistic magnetic structure of parent phase of iron-based superconductors\cite{PRB83.233205,CPL28.086104,PRL107.137003,PRL102.247001},
we construct seven kinds of initial magnetic structures as possible candidates,
including nonmagnetic (NM), ferromagnetic (FM), N\'{e}el antiferromagnetic (NAFM),
striped AFM(SAFM), blocked checkerboard AFM (BCAFM), bi-collinear AFM (BAFM) and
zigzag AFM (ZAFM) states. The last three configurations are shown in Fig.
\ref{fig:AFMstruct} (a)-(c), respectively.

\section{Numerical Results and Discussions}
Within the GGA scheme, we first find that FM, NAFM and SAFM states are unstable
and they all converge to the NM state without spontaneous magnetization.
The BCAFM order is stable. However, both the BAFM and ZAFM orders
are unstable and the system converges to a particular AFM order, corresponding
to such a $\text{2}\times \text{1}\times \text{1}$ magnetic unit cell which
contains two titanium atoms without magnetic moments and two others with
antiparallel moments, as shown in Fig. \ref{fig:AFMstruct} (d).
In such a magnetic structure the AFM moments of Ti atoms are site-selective.
More importantly, we find that the BCAFM and site-selective AFM
phases are almost degenerate to the NM phase in total energy. The relative
total energies per Ti atom referencing to the NM phase and the magnetic moment
around each Ti atom are listed in Line (a) in Table I.

For these almost degenerate phases, we expect that electronic correlation would
remove the degeneracy.
Thus the GGA$+U$ calculations are performed for the different effective Coulomb
correlation $U_{eff}=U-J$ where $U$ and $J$ are the on-site Coulomb
and Hund's exchange interactions, respectively. Unlike the GGA results, the FM
and NAFM phases are stable, but have relatively high total energies.
The SAFM phase remains unstable. The BAFM and ZAFM phases
are still unstable and both converge to the site-selective AFM state.
On the other hand, the consideration of electronic correlation correction $U_{eff}$
lifts the degeneracy of the NM, BCAFM and
site-selective AFM states, and the total energy of the site-selective AFM state
becomes the lowest.
Lines (b) and (c) in Table I summarize the relative total energies and magnetic
moments for $U_{eff}=2$ and $3$ eV. Meanwhile, one notices that several bands
cross Fermi level for different $U_{eff}$ in the lowest energy state, suggesting
that the site-selective AFM phase is metallic, in agreement with the experimental
observation\cite{JPCM22.075702}.

\begin{table*}[!htbp]
\caption{Relative total energy per Ti atom and magnetic moment around each Ti atom
in different magnetic states with $U_{eff}=0$ eV (a), $2$ eV (b) and $3$ eV (c).
The bar `` - '' means that the corresponding magnetic state is unstable and will
converge to an NM state.}
\begin{tabular}{c|cccccccc}
\hline
\hline
& Magnetic structure & \ \ \ \ NM \ \ \ \ & \ \ \ \  FM  \ \ \ \ &  \ \ \ \
NAFM  \ \ \ \ & \ \ BCAFM \ \ & \ \ \ site-selective AFM \\
\hline
 (a) $U_{eff}=0$ eV \ \ \ \
 & Relative energy (meV) & $0$ & - & -  & $-0.50$ & $-0.44$ \\
 & Ti moment (${{\mu }_{B}}$) & $0$ & - &  - & $0.144$ & $0.193$ \\
\hline
\hline
 (b) $U_{eff}=2$ eV \ \ \ \
 & Relative energy (meV) & $0$ & $-0.92$ & $-1.67$ & $-12.13$ & $-12.50$ \\
 & Ti moment (${{\mu }_{B}}$) & $0$ & $0.370$ & $0.071$ & $0.384$ & $0.538$ \\
\hline
\hline
 (c) $U_{eff}=3$ eV \ \ \ \
 & Relative energy (meV) & $0$ & $-13.74$ & $-14.10$ & $-25.31$ & $-28.19$ \\
 & Ti moment (${{\mu }_{B}}$) & $0$ & $0.516$ & $0.053$ & $0.490$ & $0.684$ \\
\hline
\end{tabular}
\end{table*}

\begin{figure}[tp]
{\epsfig{figure=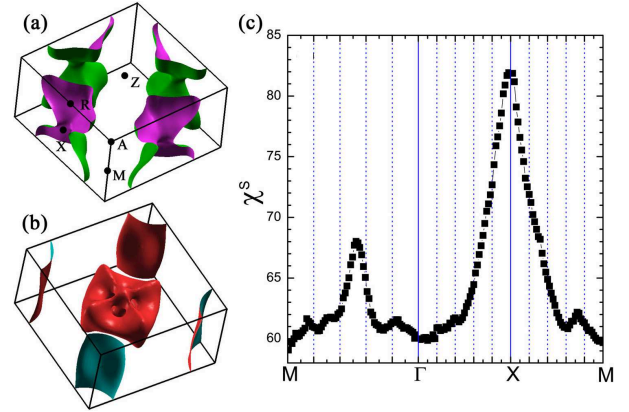,width=8cm,angle=0}}
\caption{(Color online) Fermi surfaces and spin susceptibility of BaTi$_{2}$As$_{2}$O
in NM state within the GGA sheme: (a) hole-type Fermi surface sheets,
(b) electron-type Fermi surface sheets and (c) the real part of ${{\chi }^{S}}$
with arbitrary unit.}
\label{fig:FSsus}
\end{figure}

To further explore the underlying mechanism of the phase transition at $T_p=200$ K,
we have analyzed the electronic structures of the NM which corresponds to
the high-symmetry phase at high temperature. The hole-type and electron-type Fermi
surfaces without electronic correlation correction are illustrated in
Fig. \ref{fig:FSsus} (a) and (b). The electron-type Fermi surface sheets around
the corner of Brillouin zone ($M$ point) can almost perfectly coincide with each other
after the translation along the diagonal direction. It reveals a Fermi
surface nesting with the nesting vector ${{\mathbf{Q}}_{F1}}\approx\left(
0.24\frac{2\pi }{a},\ \pm 0.24\frac{2\pi }{a},\ 0 \right)$. Besides, there is
an imperfect nesting between electron-type and hole-type Fermi surface sheets along
$\left( 1,\  0,\ 0 \right)$ or $\left( 0,\ 1,\ 0 \right)$, the corresponding
nesting vector is ${{\mathbf{Q}}_{F2}}=\left( \frac{\pi }{a},\ 0,\ 0 \right)$ or
$\left( 0,\ \frac{\pi }{a},\ 0 \right)$. Such two Fermi surface nestings would
likely lead to an SDW instability in association with the observed anomalies in
the experiment. Nevertheless, we need to perform further analysis to uniquely
determine the realistic Fermi surface nesting vector that the SDW oscillates
as $\mathbf{M}\cdot \cos \left( {{\mathbf{Q}}_{F}}\cdot \mathbf{R} \right)$ on
the Ti-Ti square lattice in BaTi$_{2}$As$_{2}$O.

In order to clarify which one of the above two nesting vectors dominates the
ground-state magnetic structure, we have studied the spin susceptibility
\begin{align}
 {{\chi }^{S}}\left( \mathbf{q} \right)=-\frac{1}{N}\sum\limits_{k,m,n}{
 \frac{f\left( {{\varepsilon }_{m}}\left( \mathbf{k}+\mathbf{q} \right)
 \right)-f\left( {{\varepsilon }_{n}}\left( \mathbf{k} \right)
 \right)}{{{\varepsilon }_{m}}\left( \mathbf{k}+\mathbf{q} \right)-
 {{\varepsilon }_{n}}\left( \mathbf{k} \right)+i\eta }}.
\end{align}
The real part of ${{\chi }^{S}}$ along $M-\Gamma-X-M$ is shown in
Fig. \ref{fig:FSsus} (c). There are two peaks almost at halfway of the $\Gamma -M$
length and at the $X$ point, respectively. They are in well agreement with
the above two nesting vectors. At the same time, we notice that the second peak,
corresponding to the wavevector $\left( \frac{\pi }{a},\ 0,\ 0 \right)$ or
$\left( \ 0,\frac{\pi }{a},\ 0 \right)$, is dominant. Therefore, it can be
expected that ${{\mathbf{Q}}_{F2}}$ may determine the ground-state magnetic
structure.

Moreover, we have studied the electronic structures of the
NM phase by using GGA$+U$ approach. When the effective Coulomb correlation
$U_{eff}$ on Ti atom increases from $2$ to $3$ eV, there is no significant qualitative
change, including the Fermi surface nesting and spin susceptibility. It is worthy
of noticing that the site-selective AFM order, which corresponds to the lowest
energy state in the GGA$+U$ calculations, just meets the Fermi surface nesting
vector ${{\mathbf{Q}}_{F2}}$ obtained in the NM state. This further confirms that the
ground state of BaTi$_{2}$As$_{2}$O is the site-selective AFM phase.
And considering the electronic correlation strength
in other iron pnictides, we think that $U_{eff}=2$ eV is a proper Coulomb
correlation. For such a correlation the energy of the site-selective ground state is 
lower about $12.5$ meV/Ti
than that of the NM state. Such an energy difference is in good agreement with the
phase transition temperature $200$ K.

\begin{figure}[tp]
{\epsfig{figure=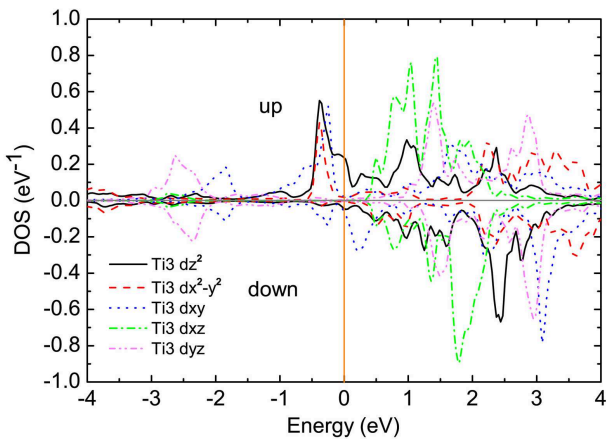,width=8.5cm,angle=0}}
\caption{(Color online) Density of states (DOS) projected into the Ti$3-3d$
orbitals for up and down spins within the GGA+U scheme (U$_{eff}$=2 eV).
Note that all DOS are depicted in the magnetic unit cell of site-selective AFM
state with $U_{eff}=2$ eV. $x$, $y$ and $z$ are directed along the $a$, $b$ and $c$
axis, respectively.}
\label{fig:DOS}
\end{figure}

In Fig.3 we have presented the electronic structures of the site-selective AFM ground state
in the correlation correction with $U_{eff}=2$ eV. The greatest
contribution to the density of states (DOS) at Fermi level is from Ti $3d$ orbitals.
It also confirms that Ti$1$ and Ti$2$ are NM, while Ti$3$ and Ti$4$ have antiparallel
magnetic moments, about $0.54 {{\mu }_{B}}$. For Ti$3$ and Ti$4$ atoms, their partial
DOS for up and down spins are almost equal except from $-1.0$ eV to $E_{F}$, as
illustrated in Fig. \ref{fig:DOS}. Such a splitting results in the formation of magnetic
moments. Nearly localized $3d_{x^{2}-y^{2}}$ and $3d_{xy}$
electrons contribute to a part of magnetic moment, and itinerant $3d_{z^{2}}$ electron
also plays a key role in the formation of the magnetic order.

\section{Conclusion}
In summary, our detailed investigations on the electronic structure and spin
susceptibility of BaTi$_{2}$As$_{2}$O have shown that both the Fermi surface
nesting and Coulomb correlation drive the site-selective AFM metallic phase as the
SDW-type ground state and the CDW order is excluded.
We expect that the existence of the site-selective AFM order with a weak magnetic
moment could be measured in future nuclear magnetic resonance experiment.
\\

This work was supported by the NSF of China under Grant No.11104274 and 11204311,
Knowledge Innovation Program of Chinese Academy of Sciences, and Director Grants
of CASHIPS. Numerical calculations were performed in Center for Computational
Science of CASHIPS.
\\

\end{document}